\documentclass[a4paper]{article}
\usepackage[a4paper,top=3cm,bottom=2cm,left=3cm,right=3cm,marginparwidth=1.75cm]{geometry}

\usepackage{amsmath}
\usepackage{appendix}
\usepackage{graphicx}
\usepackage[colorinlistoftodos]{todonotes}
\usepackage{amsthm}
\usepackage{mathrsfs}
\usepackage{amssymb}
\usepackage{accents}
\usepackage{extarrows}
\usepackage{makecell}
\usepackage{bm}
\usepackage{authblk}
\usepackage{url}
\usepackage{hyperref}
\usepackage{cite}
\usepackage{eufrak}
\hypersetup{colorlinks,
            citecolor=black,
            linkcolor=black,
            urlcolor=green,
            pdftex}

\usepackage[english]{babel}
\usepackage[utf8x]{inputenc}
\usepackage[T1]{fontenc}

\title{Hamiltonian structure and connection-dynamics of Weyl gravity}
\author[1]{Qian Chen \footnote{201521140012@mail.bnu.edu.cn}}
\author[1,2]{Yongge Ma \footnote{mayg@bnu.edu.cn}\thanks{corresponding author}}

\affil[1]{Department of Physics, Beijing Normal University, Beijing 100875, China}
\affil[2]{Institute for Gravitation and the Cosmos \& Physics Department, The Pennsylvania State University, University Park, PA 16802 U.S.A.}

\date{}

\begin{document}

\maketitle

\begin{abstract}
  A crucial property of Weyl gravity is its conformal invariance. It is shown how this gauge symmetry is exactly reflected by the two constraints in the Hamiltonian framework. Since the spatial 3-metric is one of the configuration variables, the phase space of Weyl gravity can be extended to include internal gauge freedom by triad formalism. Moreover, by canonical transformations, we obtain two new Hamiltonian formulations of Weyl gravity with an SU(2) connection as one of its configuration variables. The connection-dynamical formalisms lay the foundation to quantize Weyl gravity nonperturbatively by applying the method of loop quantum gravity. In one of the formulations, the so-called Immirzi parameter ambiguity in loop quantum gravity is avoided by the conformal invariance.\\

\end{abstract}

\section{Introduction}
Modified gravity theories have received increasingly attention due to motivations coming from cosmology, astrophysics as well as quantum gravity. One of the most interesting theories of modified gravity is the Weyl gravity \cite{Weyl1918}, whose action is defined by the square of the Weyl
tensor $C_{\mu\nu\rho\sigma}$ as
\begin{equation} \label{action}
I=-\frac14\int d^4x C_{\mu\nu\rho\sigma}C^{\mu\nu\rho\sigma}\sqrt{-g} ,
\end{equation}
where we consider 4-dimensional Lorentzian spacetimes and use the geometrical unit system, $g$ denotes the determinant of the spacetime metric $g_{\mu\nu}$. Besides the diffeomorphism invariance, the other intriguing property of this theory is its invariance under the local conformal transformation of the spacetime metric, $g_{\mu\nu}\to\Omega^2 g_{\mu\nu}$. As a higher-order derivative theory of gravity, it is argued that its perturbative quantization is renormalizable \cite{Stelle1977}. Moreover, Weyl gravity is closely related to supergravity \cite{Bergshoeff1981,deWit1983} and it also emerges from the twistor string theory \cite{Berkovits2004}.
Furthermore, Weyl gravity is also closely related to Einstein's general relativity (GR). This fact can be seen by comparing the equations of motion of the two theories \cite{Mannheim2012}. It is also argued that Weyl gravity could be employed to account for the dark matter problem (see \cite{Mannheim2012} and references therein).

The variation of action (\ref{action}) leads to the following Bach equation \cite{Bach1921}
\begin{equation}
2\nabla_{\beta}\nabla_{\alpha}C^{\alpha\mu\nu\beta}+C^{\alpha\mu\nu\beta}R_{\alpha\beta}=0 .
\end{equation}
Alternatively, action (\ref{action}) can also be written as
\begin{equation} \label{R plus G}
I=\int 2(R_{\mu\nu}R^{\mu\nu}-\frac13 R^2)\sqrt{-g}d^4x+\int G\sqrt{-g}d^4x ,
\end{equation}
where the integral of the term $G$ will give the Gauss-Bonnet-Chern topological invariant \cite{Fenchel}. Hence this term does not contribute to the equations of motion. The variation of the first term in action (\ref{R plus G}) leads to the following equivalent form of Bach equation \cite{Mannheim2012}
\begin{equation*} \begin{split}
    0=&\frac12g^{\mu\nu}R^{;\alpha}_{\phantom{;\alpha};\alpha}+R^{\mu\nu;\alpha}_{\phantom{\mu\nu;\alpha};\alpha}
    -R^{\mu\alpha;\nu}_{\phantom{\mu\alpha;\nu};\alpha}
    -R^{\nu\alpha;\mu}_{\phantom{\nu\alpha;\mu};\alpha}-2R^{\mu\alpha}R^{\nu}_{\phantom{\nu}\alpha} \\
    &+\frac12g^{\mu\nu}R_{\alpha\beta}R^{\alpha\beta}-\frac23g^{\mu\nu}R^{;\alpha}_{\phantom{;\alpha};\alpha}+\frac23R^{;\mu;\nu}
    +\frac23RR^{\mu\nu}-\frac16g^{\mu\nu}R^2 .
\end{split} \end{equation*}
Then it is straightforward to see that the solution of vacuum Einstein equation, $R_{\mu\nu}=0$, is also a solution of vacuum Weyl gravity. Hence, the solution set of vacuum Weyl gravity contains all solutions of vacuum Einstein gravity. An interesting question is whether the different conformally equivalent classes of the solutions of Weyl gravity can be characterized by the different solutions of GR? The answer is negative. In particular, it is shown that there exist solutions to Bach equation that are not conformally equivalent to Einstein spaces \cite{Schmidt1984,Nurowski2001,Liu2013}. This fact implies richer structures in Weyl gravity than those in GR. Hence Weyl gravity may bring more interesting physical phenomena in our eye shot.

The goal of this paper is to set up a classical Hamiltonian formulation towards nonperturbative quantization of Weyl gravity. It is well known that loop quantum gravity (LQG) has been widely investigated for quantizing GR \cite{Ashtekar1991,Rovelli2004,Thiemann2007,Ashtekar2004,Ma2007} as well as scalar-tensor theories of gravity \cite{Ma2011a,Ma2011b}. One of the impressive aspects of LQG is the so-called background independence. This background-independent quantization approach relies on the key observation that classical GR and scalar-tensor gravity can be cast into the connection-dynamical formalism with the
structure group of SU(2) \cite{Ashtekar1986,Barbero1995,Ma2013}. Bases on the geometrodynamics of Weyl gravity in \cite{Tureanu2014}, this paper is devoted to establish the connection-dynamical formalism for Weyl gravity.

In section \ref{CWG}, we discuss the two conformal constraints in the Hamiltonian framework of Weyl gravity, which turn out to be generators of spatial and temporal conformal transformations respectively.
In section \ref{S:Ff}, we bring triad language into the spatial metric for the sake of going towards connection-dynamical formalism. The triad formalism has an additional constraint with respect to the rotation gauge freedom of the triad. The first-class property of the constraint algebra is unchanged as the rotation constraint is imposed. The gauge transformations generated by the constraints are analysed. In section \ref{c-d F}, we derive the connection-dynamical formalisms of Weyl gravity in two different schemes by canonical transformations from its triad formalism. The Gaussian and diffeomorphism constraints in the connection formalism are similar to those of GR coupling to matters \cite{Thiemann2007}. The so-called Immirzi parameter ambiguity can be avoided in one of the schemes. The results of this paper are summarized and remarked in the last section.

\section{Conformal constraints in canonical Weyl gravity} \label{CWG}
\subsection{Geometrodynamics}
In this subsection we briefly outline the geometrical dynamics of Weyl gravity obtained in \cite{Tureanu2014}. By a (3+1) decomposition of spacetime, one obtains the induced spatial 3-metric $h_{ab}$ and the extrinsic curvature $K_{cd}$ of the foliation hypersurface $\Sigma_t$. The action (\ref{action}) can be written as
\begin{equation} \label{3+1 action}
I=\int dt\int_{\Sigma_t} d^3x N\sqrt{h}\left(C^{abc}_{\phantom{abc}\mathbf{n}}C_{abc\mathbf{n}} -2C^{a\phantom{\mathbf{n}}b}_{\phantom{a}\mathbf{n}\phantom{b}\mathbf{n}}C_{a\mathbf{n}b\mathbf{n}}\right) ,
\end{equation}
where $h$ represents the determinant of $h_{ab}$, we have denoted $C_{abc\mathbf{n}}\equiv C_{\mu\nu\rho\sigma}h_a^{\mu}h_b^{\nu}h_c^{\rho}n^{\sigma}$ and $C_{a\mathbf{n}b\mathbf{n}}\equiv C_{\mu\rho\nu\sigma}h_a^{\mu}h_b^{\nu}n^{\rho}n^{\sigma}$ respectively, with $n^{\sigma}$ being the unit normal of $\Sigma_t$. Note that the Weyl tensor contains the derivative of the extrinsic curvature as
\begin{equation}
C_{a\mathbf{n}b\mathbf{n}}=-\frac12\left(\delta^c_a\delta^d_b-\frac13h_{ab}h^{cd}\right)\left(\pounds_n K_{cd}-R_{cd}-K_{cd}K-\frac{1}{N}D_c D_d N\right)
\end{equation}
and
\begin{equation} \label{C-abcn}
C_{abc\mathbf{n}}=2D_{[a}K_{b]c}+D_d K^d_{[a} h_{b]c}-D_{[a}K h_{b]c} ,
\end{equation}
where $N$ is the lapse function, $\pounds_n$ denotes the Lie derivative along $n^{\nu}$ and $D_a$ denotes the spatial covariant derivative compatible with $h_{ab}$. One could check that action (\ref{3+1 action}) is still invariant for conformal transformations $g_{\mu\nu}\to\Omega^2 g_{\mu\nu}$.

The 3+1 form consists of basic variables $(h_{ab}, K_{ab}, \pounds_{t}K_{ab}, N, N^a)$, where $N^a$ is the shift vector. In order to reduce this higher order derivative theory into second-order derivative one, a Lagrangian multiplier $\lambda^{ab}$ is introduced into the action as
\begin{equation} \label{H action}
I=\int dt\int_{\Sigma_t} d^3x N\sqrt{h}\left(C^{abc}_{\phantom{abc}\mathbf{n}}C_{abc\mathbf{n}} -2 C^{a\phantom{\mathbf{n}}b}_{\phantom{a}\mathbf{n}\phantom{b}\mathbf{n}}C_{a\mathbf{n}b\mathbf{n}}+\lambda^{ab}(\pounds_n h_{ab}-2K_{ab})\right).
\end{equation}
Then the basic variables are increased as $(h_{ab}, \pounds_{t}h_{ab}, K_{ab}, \pounds_{t}K_{ab}, N, N^a, \lambda^{ab})$. In Hamiltonian formulation, one obtains momentum variables conjugate to the 3-metric and extrinsic curvature respectively as
\begin{equation} \label{momenta}
   \begin{aligned}
&\pi^{cd}=\lambda^{cd}\sqrt{h} , \\
&\mathcal P^{cd}=2C^{c\phantom{\mathbf{n}}d\phantom{\mathbf{n}}}_{\phantom{c}\mathbf{n}\phantom{d}\mathbf{n}}\sqrt{h} , \\
\end{aligned}
\end{equation}
with the canonical relations
\begin{equation}
\{h_{ab}(x), \pi^{cd}(y)\}=\{K_{ab}(x), \mathcal P^{cd}(y)\}=\delta_{(a}^c\delta_{b)}^d\delta^3(x,y) .
\end{equation}
From action (\ref{H action}), one can easily derive the diffeomorphism constraint $H_a$ and Hamiltonian constraint $H_0$ as
\begin{equation} \label{D & H constraints}
   \begin{aligned}
&H_a=-2h_{ab}D_c \pi^{bc}+\mathcal P^{bc}D_a K_{bc}-2D_b (\mathcal P^{bc} K_{ac})\approx 0 , \\
&H_0 = 2\pi^{ab}K_{ab}-\frac{\mathcal P_{ab}\mathcal P^{ab}}{2\sqrt{h}}+\mathcal P^{ab}R_{ab}+\mathcal P^{ab}K_{ab}K+D_aD_b \mathcal P^{ab}-\sqrt{h} C_{abc\mathbf{n}}C^{abc\phantom{\mathbf{n}}}_{\phantom{abc}\mathbf{n}}\approx 0 , \\
\end{aligned}
\end{equation}
where $\approx$ means "equal on the constraint surface".
Moreover, one obtains the following two conformal constraints due to the traceless of $\mathcal P^{cd}$ and its consistency condition:
\begin{equation} \label{conformal constraints}
   \begin{aligned}
&\mathcal P=h_{ab}\mathcal P^{ab}\approx 0,\\
&\mathcal Q=2h_{ab}\pi^{ab}+K_{ab}\mathcal P^{ab}\approx 0.\\
\end{aligned}
\end{equation}
One can check that all the constraints are of first class. Hence the physical degrees of freedom of Weyl gravity reduce to $6(=6+6-4-2)$.

\subsection{Conformal gauge transformation} \label{i c.t}
The conformal invariance of action (\ref{action}) is encoded in the constraints (\ref{conformal constraints}) in the Hamiltonian formalism. In this subsection we will show how to generate spacetime conformal transformations by those constraints. In order to become functions on the phase space, the two constraints (\ref{conformal constraints}) should be smeared over suitable test fields $\omega_{\ell}(x)$ and $\omega_{\perp}(x)$ as \begin{equation} \begin{aligned}
\mathcal P(\omega_{\perp})=\int_{\Sigma_t} d^3x\mathcal P\omega_{\perp}, \\
\mathcal Q(\omega_{\ell})=\int_{\Sigma_t}d^3x\mathcal Q\omega_{\ell}.
\end{aligned}\end{equation}
Then it is straightforward to get
\begin{equation} \label{c.t by Q}
   \begin{aligned}
\{h_{ab}, \mathcal Q(\omega_{\ell})\}&=2\omega_{\ell} h_{ab}, \\
\{\pi^{ab}, \mathcal Q(\omega_{\ell})\}&=-2\omega_{\ell} \pi^{ab}, \\
\{K_{ab}, \mathcal Q(\omega_{\ell})\}&=\omega_{\ell} K_{ab}, \\
\{\mathcal P^{ab}, \mathcal Q(\omega_{\ell})\}&=-\omega_{\ell} \mathcal P^{ab},
\end{aligned}
\end{equation}
and
\begin{equation} \label{c.t by P}
   \begin{aligned}
\{h_{ab}, \mathcal P(\omega_{\perp})\}&=0, \\
\{\pi^{ab}, \mathcal P(\omega_{\perp})\}&=-\omega_{\perp} \mathcal P^{ab}, \\
\{K_{ab}, \mathcal P(\omega_{\perp})\}&=\omega_{\perp} h_{ab}, \\
\{\mathcal P^{ab}, \mathcal P(\omega_{\perp})\}&=0,
\end{aligned}
\end{equation}
respectively. Note that the infinitesimal transformations of $\pi^{ab}$ in (\ref{c.t by Q}) and (\ref{c.t by P}) imply that the Lagrange multiplier $\lambda^{ab}$ introduced in action (\ref{H action}) has to be transformed as
\begin{equation}
\lambda^{ab} \to \Omega^{-5}(\lambda^{ab}-2C^{a\phantom{\mathbf{n}}b\phantom{\mathbf{n}}}_{\phantom{a}\mathbf{n}\phantom{b}\mathbf{n}}n^{\mu}\partial_{\mu}\ln\Omega)
\end{equation}
under a finite conformal transformation: $g_{\mu\nu}\to\Omega^2g_{\mu\nu}$. The finite spacetime conformal transformation induces transformations on $\Sigma_t$ as
\begin{equation} \label{4d c.t} \begin{split}
h_{ab}&\to\Omega^2h_{ab} ,\\
K_{ab}&\to\Omega K_{ab}+h_{ab}n^{\mu}\partial_{\mu}\Omega , \\
\mathcal P^{ab} &\to \Omega^{-1}\mathcal P^{ab} ,
\end{split} \end{equation}
where $n_{\mu}\to\Omega n_{\mu}$ and $K_{ab}=\frac12\pounds_n h_{ab}$ are used. The relation between the conformal factor $\Omega$ and the test fields $\omega_{\ell}$ and $\omega_{\perp}$ can be explored, if the transformations (\ref{c.t by Q}) and (\ref{c.t by P}) generated by constraints $Q(\omega_{\ell})$ and $P(\omega_{\perp})$ contribute the infinitesimal version of (\ref{4d c.t}).

Note that finite conformal transformations on the phase space can be constructed by the exponential maps of the Hamiltonian vector fields dual to functions $\mathcal Q(\omega_{\ell})$ and $\mathcal P (\omega_{\perp})$.
However, (\ref{c.t by Q}) and (\ref{c.t by P}) imply that the action order of the exponential maps $\exp[X_{\mathcal Q(\omega_{\ell})}]$ and $\exp[X_{\mathcal P(\omega_{\perp})}]$ will effect the resulted transformation of the extrinsic curvature $K_{ab}$. A straightforward calculation gives
\begin{equation} \label{PQ order} \begin{aligned}
\exp[X_{\mathcal P(\omega_{\perp})}]\exp[X_{\mathcal Q(\omega_{\ell})}]\circ K_{ab}=&\sum^{\infty}_{k=0}\frac{1}{k!}\left\{\left(\sum^{\infty}_{n=0}\frac{1}{n!}\{K_{ab},\mathcal Q(\omega_{\ell})\}_{(n)}\right),\mathcal P(\omega_{\perp})\right\}_{(k)} \\
=&\bar\Omega K_{ab}+\omega_{\perp}\bar\Omega h_{ab},
\end{aligned} \end{equation}
where $\bar\Omega\equiv \sum^{\infty}_{n}\frac{1}{n!}\omega_{\ell}^n=e^{\omega_{\ell}}$, and the suffix on the Poisson bracket denotes the iteration: $\{K_{ab},\mathcal Q(\omega_{\ell})\}_{(n+1)}=\{\{K_{ab},\mathcal Q(\omega_{\ell})\}_{(n)},\mathcal Q(\omega_{\ell})\}$. On the other hand, another order of action gives
\begin{equation} \label{QP order}
\exp[X_{\mathcal Q(\omega_{\ell})}]\exp[X_{\mathcal P(\omega_{\perp})}]\circ K_{ab}=\bar\Omega K_{ab}+\omega_{\perp}\bar\Omega^2 h_{ab}.
\end{equation}
Therefore it is obvious that
\begin{equation} \label{nc c.t}
\exp[X_{\mathcal P(\omega_{\perp})}]\exp[X_{\mathcal Q(\omega_{\ell})}]\neq \exp[X_{\mathcal Q(\omega_{\ell})}]\exp[X_{\mathcal P(\omega_{\perp})}].
\end{equation}
This noncommutative property can be understood as follows. The Poisson algebra
\begin{equation}
\{\mathcal P(\omega_{\perp}), \mathcal Q(\omega_{\ell})\}=\mathcal P(\omega_{\ell}\cdot\omega_{\perp}),
\end{equation}
together with Jacobi identity
\begin{equation}
\{\{K_{ab}, \mathcal Q(\omega_{\ell})\}, \mathcal P(\omega_{\perp}) \} + \{\{\mathcal Q(\omega_{\ell}), \mathcal P(\omega_{\perp}) \}, K_{ab}\} + \{ \{\mathcal P(\omega_{\perp}), K_{ab} \}, \mathcal Q(\omega_{\ell})\}=0,
\end{equation}
gives
\begin{equation}
\{\{K_{ab}, \mathcal Q(\omega_{\ell})\}, \mathcal P(\omega_{\perp}) \} + \omega_{\ell}\omega_{\perp}h_{ab} =\{ \{K_{ab}, \mathcal P(\omega_{\perp}) \}, \mathcal Q(\omega_{\ell})\},
\end{equation}
which implies (\ref{nc c.t}). However, there is no such a problem for the spatial metric $h_{ab}$ due to $\{h_{ab}, \mathcal P(\omega_{\perp})\}=0$.

Suppose that the Hamiltonian vector field of the linear combination,
\begin{equation}
\mathcal C(\omega_{\ell},\omega_{\perp})=\mathcal Q(\omega_{\ell})+\mathcal P(\omega_{\perp}),
\end{equation}
generates a spacetime conformal transformation (\ref{4d c.t}). By imploying the Lie product formula in Lie group theory,
\begin{equation} \begin{split}
\exp[X_{\mathcal Q(\omega_{\ell}) }+X_{\mathcal P(\omega_{\perp})}] =& \lim_{n\to \infty}\left(\exp\left[ X_{\frac{1}{n}\mathcal Q(\omega_{\ell}) }\right] \exp\left[ X_{\frac{1}{n}\mathcal P(\omega_{\perp})}\right]\right)^n \\
=&\lim_{n\to \infty}\left(\exp\left[ X_{\frac{1}{n}\mathcal P(\omega_{\perp}) }\right] \exp\left[ X_{\frac{1}{n}\mathcal Q(\omega_{\ell})}\right]\right)^n,
\end{split} \end{equation}
the above order ambiguity can be avoided. A straightforward calculation (see Appendix \ref{C=Q+P}) shows that the test fields are related to the conformal factor by
\begin{align}
& \omega_{\ell}=\ln\Omega\big|_{\Sigma_t}, \label{omega_ell}\\
& \omega_{\perp}=\frac{(\ln \Omega) n^{\mu}\partial_{\mu}\Omega}{\Omega^2-\Omega}\Big|_{\Sigma_t}. \label{omega_perp}
\end{align}

\section{Triad formalism} \label{S:Ff}
\subsection{Canonical variables in extended phase space}
In this subsection we will extend the phase space of Weyl gravity coordinatized by $(h_{ab}, \pi^{cd}; K_{ab}, \mathcal P^{cd})$ to triad formalism in order to bring some internal gauge degrees of freedom into the theory. Let $e^a_i (i=1,2,3)$ be any triad on $\Sigma_t$ such that $h^{ab}=e^a_i e^b_j \delta^{ij}$. The denstized triad is defined as $E^a_i:=\sqrt{h}e^a_i$. We denote the inverse of $E^a_i$ by $E^j_a$ and the determinant of $E^a_i$ by $E$. Suppose $\pi^j_b$ is the variable conjugate to $E^a_i$. We equip the extended phase space coordinatized by $(\pi^i_a, E^b_j; K_{ij}, \mathcal P^{kl})$ with symplectic structure defined by
\begin{equation} \label{symplectic 1st} \begin{aligned}
&\{\pi_a^i(x),E_j^b(y)\}=\delta_a^b\delta^i_j\delta^3(x,y), \\
&\{K_{ij}(x),\mathcal P^{kl}(y)\}=\delta_{(i}^k\delta_{j)}^l\delta^3(x,y),
\end{aligned} \end{equation}
and
\begin{equation} \label{symplectic 2nd}
\{\pi^k_a(x),K^{ij}(y)\}=\{\pi^i_a(x),\mathcal P^{kl}(y)\}=\{E^b_k(x),K^{ij}(y)\}=\{E^b_i(x),\mathcal P^{kl}(y)\}=0.
\end{equation}
Note that the canonical variables $\pi^i_a(x)$ and $E^b_j(y)$ have $9$ degrees of freedom respectively, while $K_{ij}(x)$ and $\mathcal P^{kl}(y)$ have $6$ respectively. The new variables are related to the original variables by
\begin{equation} \label{Frame hKP} \begin{aligned}
h_{ab}=\delta_{ij}E^i_a E^j_b E ,&\quad \pi^{cd}=\text{to be determined}, \\
K_{ab}=K_{ij}E^i_a E^j_b E ,&\quad \mathcal P^{cd}=E^{-1}\mathcal P^{kl}E^c_k E^d_l.
\end{aligned} \end{equation}
Note that by contracting with the triad, the canonical variables $K_{ab}$ and $\mathcal P^{cd}$ can be expressed as internal tensors $K_{ij}$ and $\mathcal P^{kl}$. So the key issue is to find the expression of $\pi^{cd}$ in terms of new variables. Let $\pi^{cd}=\pi^{cd}(\pi^j_b,E^a_i,K_{ij},P^{kl})$. We can solve it from the following equations with respect to the symplectic structure (\ref{symplectic 1st}) and (\ref{symplectic 2nd}),
\begin{equation} \label{equation hKP} \begin{aligned}
&\{h_{ab}(x),\pi^{cd}(y)\}=-\int_{\Sigma_t}\frac{\delta h_{ab}(x)}{\delta E^f_i(z)}\frac{\delta \pi^{cd}(y)}{\delta \pi_f^i(z)}d^3z=\delta^c_{(a}\delta^d_{b)}\delta^3(x,y), \\
&\{K_{ab}(x),\pi^{cd}(y)\}=\int_{\Sigma_t}\left(-\frac{\delta K_{ab}(x)}{\delta E^f_i(z)}\frac{\delta \pi^{cd}(y)}{\delta \pi_f^i(z)}+\frac{\delta K_{ab}(x)}{\delta K_{ij}(z)}\frac{\delta \pi^{cd}(y)}{\delta \mathcal P^{ij}(z)}\right)d^3z=0, \\
&\{\mathcal P^{ab}(x),\pi^{cd}(y)\}=\int_{\Sigma_t}\left(-\frac{\delta \mathcal P^{ab}(x)}{\delta E^f_i(z)}\frac{\delta \pi^{cd}(y)}{\delta \pi_f^i(z)}-\frac{\delta \mathcal P^{ab}(x)}{\delta \mathcal P^{ij}(z)}\frac{\delta \pi^{cd}(y)}{\delta K_{ij}(z)}\right)d^3z=0. \\
\end{aligned}  \end{equation}
Let $\pi^{cd}\equiv\bar\pi^{cd}-U^{cd}$, where
\begin{equation}
\bar\pi^{cd}=\frac{1}{2E}(E^{(c}_k E^{d)}_l \pi^l_f E^f_k - E^c_k E^d_k \pi^l_f E^f_l)
\end{equation}
and $U^{cd}=U^{cd}(E^a_i,K_{ij},\mathcal P^{ij})$. Then the first equation in (\ref{equation hKP}) is satisfied automatically, while the second and third equations in (\ref{equation hKP}) give
\begin{equation}
U^{cd}=E^{-1}K_{il}\mathcal P^{lj}E^{(c}_i E^{d)}_j.
\end{equation}
Hence we recover $\pi^{cd}$ in extended phase space as
\begin{equation} \label{pi}
\pi^{cd}=\frac{1}{2E}\left(E^{(c}_i E^{d)}_j\pi^j_f E^f_i-E^c_i E^d_i \pi^k_f E^f_k\right)-\frac{1}{E}K_{il}\mathcal P^{lj}E^{(c}_i E^{d)}_j.
\end{equation}
By a tedious calculation, the Poisson bracket between two $\pi^{ab}$ reads
\begin{equation} \label{G,G}
\{\pi^{ab}(x), \pi^{cd}(y)\}=\frac{1}{16}(h^{ac}G^{db}+h^{bc}G^{da}+h^{ad}G^{cb}+h^{bd}G^{ca})(y)\delta^3(x,y),
\end{equation}
where $G^{ab}=E^{-1}E^a_i E^b_j G_{ij}$ with $G_{ij}\equiv 2\pi_{c[i}E^c_{j]}+4K^{\phantom{i}l}_{[i}\mathcal P_{\phantom{l}j]}^l$. Note that on the extended phase space $G_{ij}$ generates exactly the internal SO(3) rotations of the new variables, which keep the original variables $(h_{ab},\pi^{cd};K_{ab},\mathcal P^{cd})$ invariant. Hence to go back to the original phase space, we need to impose the ``rotation'' constraint
\begin{equation} \label{rotation}
G(\Lambda):=\frac12\int_{\Sigma_t} d^3x G_{ij}\Lambda^{ji}\approx0
\end{equation}
on the extended phase space, where $\Lambda^{ij}$ is an arbitrary internal anti-symmetric tensor-valued test function. In addition, the functions $G(\Lambda)$ constitute a closed constraint algebra as
\begin{equation} \label{rotation structure}
\{G(\Lambda), G(\Lambda')\}=G([\Lambda,\Lambda']).
\end{equation}
It is easy to check that
\begin{equation} \begin{split} \label{h,pi,K,P invariant G}
&\{G(\Lambda), h_{ab}(x)\}=0, \\
&\{G(\Lambda), \pi^{cd}(x)\}=0, \\
&\{G(\Lambda), K_{ab}(x)\}=0,\\
&\{G(\Lambda), \mathcal P^{cd}(x)\}=0. \\
\end{split} \end{equation}

\subsection{Triad formalism as a first-class system}
We want to show that all previous constraints together with the rotation constraints on the extended phase space constitute a first-class constrained system. Note that except for $G(\Lambda)$, all other constraints can be obtained by naive substitution of $h_{ab},\ \pi^{cd},\ K_{ab}$ and $\mathcal P^{cd}$ in (\ref{D & H constraints}) and (\ref{conformal constraints}) with (\ref{Frame hKP}) and (\ref{pi}), which denote as $\mathcal P',\ \mathcal Q',\ H'_a$ and $H'_0$ respectively. Since the expressions of $\mathcal P', \mathcal Q', H'_a$ and $H'_0$ may contain the rotation constraint which can be neglected on the constraint surface, one usually use some alternative expressions of those constraints without the terms containing the rotation constraint. We denote $\mathcal P\equiv\mathcal P'+Z_{\mathcal P},\ \mathcal Q\equiv\mathcal Q'+Z_{\mathcal Q},\ H_a\equiv H'_a+Z_{a}$, and $H_0\equiv H'_0+Z_0$, where $Z_{\mathcal P},\ Z_{\mathcal Q},\  Z_{a}$ and $Z_0$ vanish on the constraint surface of the rotation constraint. Since $\mathcal P',\ \mathcal Q',\ H_a'$ and $H_0'$ are defined in terms of (\ref{Frame hKP}) and (\ref{pi}), (\ref{h,pi,K,P invariant G}) ensures that $\mathcal P',\ \mathcal Q',\ H_a'$ and $H_0'$ are invariant under the internal rotation generated by $G(\Lambda)$. Together with (\ref{rotation structure}), we conclude that
\[
\{G,\mathcal P\},\{G,\mathcal Q\},\{G,H_a\},\{G,H_0\}\propto G \approx 0.
\]
Thus $G$ form an ideal of the constraint algebra. Since (\ref{D & H constraints}) and (\ref{conformal constraints}) are indeed first-class, we have shown that $\mathcal P, \mathcal Q, H_a, H_0$ together with $G_{ij}$ are also first-class in extended phase space. Since the constraint algebra in the original phase space is known \cite{Tureanu2014}, one can use the symplectic reduction formulas (\ref{equation hKP}) and (\ref{G,G}) to derive the constraint algebra in extended phase space. For instance, let $H'_{0}(\xi)\equiv\int_{\Sigma_t} \xi H'_0  d^3x$ and $H'_{0}(\eta)\equiv\int_{\Sigma_t} \eta H'_0 d^3x$ be the smeared Hamiltonian constraints. To calculate $\{H'_0(\xi), H'_0(\eta)\}$, we can first calculate
\begin{equation} \label{H0,H0. Ff}\begin{split}
\{H'_0(\xi), H'_0(\eta)\}=&\int_{\Sigma_t}\left(\frac{\delta H'_0(\xi)}{\delta\pi^i_a(x)}\frac{\delta H'_0(\eta)}{\delta E^a_i(x)}+\frac{\delta H'_0(\xi)}{\delta K_{ij}(x)}\frac{\delta H'_0(\eta)}{\delta\mathcal P^{ij}(x)}
-(\xi\leftrightarrow\eta)\right)d^3x \\
=&\{\bar H_0(\xi),\bar H_0(\eta)\}\vert_{\Gamma_0}+\int_{\Sigma_t}d^3 x\int_{\Sigma_t}\frac{\delta \bar H_0(\xi)}{\delta \pi^{ab}(x)}\frac{\delta \bar H_0(\eta)}{\delta \pi^{cd}(y)}\{\pi^{ab}(x), \pi^{cd}(y)\} d^3 y, \\
\end{split} \end{equation}
where $\bar H_0=\bar H_0(h_{ab},\pi^{cd},K_{ab},\mathcal P^{cd})$ is the Hamiltonian constraint coordinatized by $(h_{ab},\pi^{cd};K_{ab},\mathcal P^{cd})$, and $\{\bar H_0(\xi),\bar H_0(\eta)\}\vert_{\Gamma_0}$ takes the same result as that of the original constraint algebra. Then we substitute all functions of $(h_{ab},\pi^{cd};K_{ab},\mathcal P^{cd})$ by functions of $(\pi^i_a,E^b_j;K_{ij},\mathcal P^{kl})$. Thus we obtain the constraint algebra in extended phase space by naive substitution as
\begin{equation} \begin{split}
&\{H'_0,H'_0\}\propto H'_a\oplus\mathcal P'\oplus G,\quad \{H'_a,H'_b\}\propto H'_c\oplus G,\quad\{H'_0,H'_a\}\propto H'_0\oplus G, \\
&\{\mathcal P',H'_0\}\propto \mathcal P'\oplus\mathcal Q',\quad\{\mathcal Q',H'_0\}\propto \mathcal P'\oplus H'_0\oplus G, \quad\{\mathcal Q',H'_a\}\propto \mathcal Q'\oplus G, \\
&\{\mathcal P',H'_a\}\propto \mathcal P', \quad\{\mathcal P',\mathcal Q'\}\propto \mathcal P'.
\end{split} \end{equation}
Then it is straightforward to calculate the algebra for the constraints with $G$ linear combination as
\begin{equation} \begin{split}
\{H_0,H_0\}&=\{H'_0+Z_0,H'_0+Z_0\}=\{H'_0,H'_0\}+\{Z_0,Z_0\},  \\
\{H_a,H_b\}&=\{H'_a+Z_a,H'_b+Z_b\}=\{H'_a,H'_b\}+\{Z_a,Z_b\}, \\
\{H_0,H_a\}&=\{H'_0+Z_0,H'_a+Z_a\}=\{H'_0,H'_a\}+\{Z_0,Z_a\}, \\
&\cdots
\end{split} \end{equation}
Since the constraints form a first-class system in extended phase space, the physical degrees of freedom of Weyl gravity can also be read as $6=9+6-3-1-2-3$.

\subsection{Conformal, diffeomorphism and rotation constraints in extended phase space}
The naive substitution of the conformal constraints (\ref{conformal constraints}) in terms of new variables reads
\begin{equation} \label{c.c in Ff} \begin{aligned}
\mathcal P'=\mathcal P=&\mathcal \delta_{ij} P^{ij}\approx 0, \\
\mathcal Q'=\mathcal Q=&-(2\pi^i_a E^a_i+K_{ij}\mathcal P^{ij})\approx 0. \\
\end{aligned} \end{equation}
It is easy to check that they Poisson commute with $G(\Lambda)$,
\begin{equation}
\{G(\Lambda),\mathcal P(\omega_{\perp})\}=\{G(\Lambda),\mathcal Q(\omega_{\ell)} \}=0,
\end{equation}
where we omitted the ``primes''. $\mathcal Q(\omega_{\ell})$ and $\mathcal P(\omega_{\perp})$ still generate conformal transformations. Note that the minus sign in the expression of $\mathcal Q$ arises from the fact that in the new coordinates we employed the densitized triad $E^b_j$ as the momentum variable conjugate to $\pi^i_a$.

The naive substitution of the diffeomorphism constraint in (\ref{D & H constraints}) reads
\begin{equation}
H'_a=E^b_i D_a\pi^i_b-D_b(\pi^i_a E^b_i) +\mathcal P^{ij}D_a K_{ij}+\frac12(G_{ij}E^b_j D_a E^i_b - D_b(G_{ij}E^i_a E^b_j))\approx 0.
\end{equation}
By removing the terms containing the rotation constraint, we obtain
\begin{equation} \label{triad diff. c.}
H_a=E^b_i D_a\pi^i_b-D_b(\pi^i_a E^b_i) +\mathcal P^{ij}D_a K_{ij}\approx 0.
\end{equation}
It turns out that it is $H_a$ rather than $H'_a$ generates the spatial diffeomorphisms of the new variables, since the smeared version of $H_a$ takes the form
\begin{equation} \begin{split}
H_a(\xi^a)=&\int_{\Sigma_t}d^3x\xi^a\left(E^b_i D_a\pi^i_b-D_b(\pi^i_a E^b_i) +\mathcal P^{ij}D_a K_{ij}\right) \\
=&\int_{\Sigma_t}d^3x\left(E^a_i \pounds_{\xi}\pi^i_a +\mathcal P^{ij}\pounds_{\xi} K_{ij}\right),
\end{split} \end{equation}
where $\xi^a$ is any test vector field on $\Sigma_t$ satisfying suitable boundary condition.

The Poisson bracket between two rotation constraints can be calculated as
\begin{equation}
\{G(\Lambda),G(\Lambda')\}=G([\Lambda,\Lambda']).
\end{equation}
It is easy to see that the canonical transformations generated by $G(\Lambda)$ on $(\pi^i_a, E^b_j)$ are exactly the internal rotation as in GR \cite{Ashtekar1991,Thiemann2007}. $G(\Lambda)$ also generates internal rotations on $(K_{ij},\mathcal P^{kl})$ as
\begin{equation} \label{G generating} \begin{aligned}
\{K_{ij}(x), G(\Lambda)\}&=\Lambda^{il}K_{lj}(x)+\Lambda^{jl}K_{il}(x)=[\Lambda,K]^i_{\phantom{i}j}(x), \\
\{\mathcal P^{ij}(x), G(\Lambda)\}&=\Lambda^{il} \mathcal P^{lj}(x)+\mathcal P^{il}\Lambda^{jl}(x)=[\Lambda,\mathcal P]^i_{\phantom{i}j}(x).
\end{aligned} \end{equation}

The infinitesimal conformal transforms generated by $\mathcal Q(\omega_{\ell})$ and $\mathcal P(\omega_{\perp})$ are calculated as
\begin{equation} \label{c.t by Q in F.F.}
   \begin{aligned}
\{\pi^i_a(x), \mathcal Q(\omega_{\ell})\}&=-2\omega_{\ell} \pi^i_a(x), \\
\{E^a_i(x), \mathcal Q(\omega_{\ell})\}&=2\omega_{\ell} E^a_i(x), \\
\{K_{ij}(x), \mathcal Q(\omega_{\ell})\}&=-\omega_{\ell} K_{ij}(x), \\
\{\mathcal P^{ij}(x), \mathcal Q(\omega_{\ell})\}&=\omega_{\ell} \mathcal P^{ij}(x),
\end{aligned}
\end{equation}
and
\begin{equation} \label{c.t by P in F.F.}
   \begin{aligned}
\{\pi^i_a(x), \mathcal P(\omega_{\perp})\}&=0, \\
\{E^a_i(x), \mathcal P(\omega_{\perp})\}&=0, \\
\{K_{ij}(x), \mathcal P(\omega_{\perp})\}&=\delta_{ij}\omega_{\perp}(x), \\
\{\mathcal P^{ij}(x), \mathcal P(\omega_{\perp})\}&=0,
\end{aligned}
\end{equation}
respectively. The conformal generator $\mathcal P$ only affects $K_{ij}$ and thus $U^{cd}$ part of $\pi^{cd}$.

\section{Connection-dynamical formalism} \label{c-d F}
\subsection{The first scheme}
In the triad formalism studied in last section, the configuration variable $\pi^i_a$ is a Lie algebra $\mathfrak{so}(3)$ (or $\mathfrak{su}(2)$) valued one-form. However, $\pi^i_a$ is not a connection since the rotation constraint is not the Gaussian constraint of a gauge theory. Similar to the case of GR, we can construct a $\mathfrak{su}(2)$ connection by a canonical transformation on the extended phase space as:
\begin{equation}
A^i_a=\Gamma^i_a+\gamma\pi^i_a,
\end{equation}
where $\Gamma^i_a$ is the $\mathfrak{su}(2)$ spin connection determined by $E^b_j$
\begin{equation}
\Gamma^i_a=\frac12\epsilon^{ijk}e^b_k(\partial_b e^j_a-\partial_a e^j_b + e^l_a e^c_j\partial_b e^l_c),
\end{equation}
and $\gamma$ is an arbitrary nonzero real number. We further define ${}^{(\gamma)}E^b_j=\frac{1}{\gamma}E^b_j$. Then $(A^i_a,{}^{(\gamma)}E^b_j)$ constitute a new canonical pair. Combining the rotation constraint $G^{ij}\epsilon_{ijk}\approx 0$ with the compatibility condition:
\begin{equation} \label{pre-G}
D_a E^a_i=\partial_a E^a_i+\epsilon_{ijk}\Gamma^j_a E^a_k=0,
\end{equation}
we obtained the standard Gaussian constraint:
\begin{equation}
\mathcal G_i=\partial_a {}^{(\gamma)} E^a_i + \epsilon_{ijk}A^j_a {}^{(\gamma)}E^a_k+\epsilon_{ijk}K_{jl}\mathcal P^{lk}\approx 0.
\end{equation}
Hence $A^i_a$ is an $\mathfrak{su}(2)$ connection, and the internal tensor $K_{ij}$ and $\mathcal P^{kl}$ play the role of the source of this gauge theory.

The fundamental Poisson brackets can be derived from the symplectic structure  (\ref{symplectic 1st}) and (\ref{symplectic 2nd}) as
\begin{equation} \begin{aligned} \label{c-d Poisson}
&\{A^i_a(x), {}^{(\gamma)}E^b_j(y)\}=\delta^i_j\delta^b_a\delta^3(x,y),\quad \{K_{ij}(x),\mathcal P^{kl}(y)\}=\delta^k_{(i}\delta^l_{j)}\delta^3(x,y), \\
&\{A^i_a(x), A^j_b(y)\}=\{A^k_a(x), K_{ij}(y)\}=\{A^i_a(x), P^{kl}(y)\}=0, \\
&\{{}^{(\gamma)}E^a_i(x), {}^{(\gamma)}E^b_j(y)\}=\{{}^{(\gamma)}E^a_j(x), K_{ij}(y)\}=\{{}^{(\gamma)}E^a_i(x), P^{kl}(y)\}=0.
\end{aligned} \end{equation}
Since the Gaussian constraint is a linear combination of the rotation constraint and the compatibility condition, it also contributes a closed constraint algebra:
\begin{equation}
\{\mathcal G (\Lambda), \mathcal G (\Lambda')\}=\mathcal G ([\Lambda,\Lambda']).
\end{equation}
The curvature of $A^i_a$ reads
\begin{equation} \label{curvatures}
F^i_{ab}=2\partial_{[a} A^i_{b]} + \epsilon_{ijk} A^j_a A^k_b.
\end{equation}
One can define a new covariant derivative $\mathcal D_a$ associated with connection $A^i_a$ by
\begin{equation}
\mathcal D_a V^i=\partial_a V^i+\epsilon_{ijk}A^j_a V^k.
\end{equation}
The original geometric variables can be rewritten in terms of new variables as
\begin{equation} \begin{aligned}
h_{ab}&=\gamma { }^{(\gamma)}E { }^{(\gamma)}E^i_a { }^{(\gamma)}E^i_b, \\
\pi^{cd}&=\frac{1}{2\gamma { }^{(\gamma)}E}\left[{ }^{(\gamma)}E^{(c}_j { }^{(\gamma)}E^{d)}_i (A^i_a-\Gamma^i_a) { }^{(\gamma)}E^a_j - { }^{(\gamma)}E^c_j { }^{(\gamma)}E^d_j (A^i_a-\Gamma^i_a) { }^{(\gamma)}E^a_i-2K_{il}\mathcal P^{lj}{ }^{(\gamma)}E^{(c}_i { }^{(\gamma)}E^{d)}_j\right], \\
K_{ab}&=\gamma { }^{(\gamma)}E { }^{(\gamma)}E^i_a { }^{(\gamma)}E^j_b K_{ij}, \\
\mathcal P^{cd}&=\gamma^{-1}{ }^{(\gamma)}E^{-1} { }^{(\gamma)}E^c_k { }^{(\gamma)}E^d_l\mathcal P^{kl}.
\end{aligned} \end{equation}
Then the constraints can be recast as
\begin{equation} \label{c.t. in c-d} \begin{aligned}
\mathcal G_i&=\mathcal D_a { }^{(\gamma)}E^a_i+2\epsilon_{ijk}K_{jl}\mathcal P^{lk}\approx 0, \\
\mathcal P&=\delta_{ij}\mathcal P^{ij}\approx 0, \\
\mathcal Q&=-2(A^i_a-\Gamma^i_a){ }^{(\gamma)}E^a_i-K_{ij}\mathcal P^{ij}\approx 0, \\
H_a&=F^i_{ab} { }^{(\gamma)}E^b_i+\mathcal P^{ij} \mathcal D_a K_{ij}-\gamma\pi^i_a\mathcal G_i\approx 0, \\
H_0&=\gamma^{-\frac32}\mathcal H_{A}+\gamma^{-1}\mathcal H_{B}+\mathcal H_C+\gamma^{\frac12}\mathcal H_D\approx 0,
\end{aligned} \end{equation}
where $H_a$ and $H_0$ can be derived from (\ref{triad diff. c.}) and (\ref{D & H constraints}) by naive substitution respectively, and the terms $\mathcal H_{A}$, $\mathcal H_{B}$, $\mathcal H_{C}$ and $\mathcal H_{D}$ can be expressed in term of new variables as
\begin{equation} \label{constraints in c-d 1st} \begin{aligned}
\mathcal H_A=&-\frac{1}{2\sqrt{{ }^{(\gamma)} E}}\mathcal P^{ij}\mathcal P^{ij}, \\
\mathcal H_B=&{ }^{(\gamma)} E^a_{(i} { }^{(\gamma)} E^b_{j)} { }^{(\gamma)} E^{-1}\Big[ \mathcal D_a \mathcal D_b \mathcal P^{ij}-4\epsilon^{i}_{\phantom{i}kl}{ }^{(\gamma)}\pi^k_a \mathcal D_b \mathcal P^{jl} -2\epsilon^{i}_{\phantom{i}kl}  \mathcal P^{jl}\mathcal D_b{ }^{(\gamma)}\pi^k_a +6\mathcal P^{jk}{ }^{(\gamma)}\pi^k_a { }^{(\gamma)}\pi^i_b \\
&-4\mathcal P^{ij}{ }^{(\gamma)}\pi^k_a{ }^{(\gamma)}\pi^k_b - 2\delta_{ij}\mathcal P^{kl}{ }^{(\gamma)}\pi^k_a{ }^{(\gamma)}\pi^l_b \Big]+{ }^{(\gamma)} E^{-1}\mathcal P^{ij}{ }^{(\gamma)} E^a_j{ }^{(\gamma)} E^b_k R^{ik}_{ab}, \\
\mathcal H_C=&K_{ij}{ }^{(\gamma)}\pi^i_a { }^{(\gamma)} E^a_j-3K { }^{(\gamma)}\pi^i_a { }^{(\gamma)} E^a_i -2K_{ij}K_{il}\mathcal P^{lj}, \\
\mathcal H_D=&-\sqrt{{ }^{(\gamma)} E}C_{abc\mathbf{n}}C^{abc}_{\phantom{abc}\mathbf{n}}.
\end{aligned} \end{equation}
Note that ${ }^{(\gamma)}\pi^i_a\equiv\gamma\pi^i_a=A^i_a-\Gamma^i_a$ does not depend on $\gamma$ actually, and we have made use of the conformal constraints $\mathcal Q$ and $\mathcal P$ for sake of obtaining $\mathcal H_B$ and $\mathcal H_C$. The expression of $C_{abc\mathbf{n}}C^{abc}_{\phantom{abc}\mathbf{n}}$ reads
\begin{equation}
C_{abc\mathbf{n}}C^{abc}_{\phantom{abc}\mathbf{n}}=\epsilon^{abd}\epsilon^{fgc}(D_{a}K_{bc})D_{f} K_{gd}+\epsilon^{abd}\epsilon^{fg}_{\phantom{fg}d}(D_{a}K_{bc}) D_{f} K_{g}^{\phantom{g}c},
\end{equation}
which can be rewritten in term of new variables as
\begin{equation} \begin{split}
C_{abc\mathbf{n}}C^{abc}_{\phantom{abc}\mathbf{n}}=&{ }^{(\gamma)}E^{-1}{ }^{(\gamma)}E^a_m { }^{(\gamma)}E^b_n \epsilon^{ijm}\epsilon^{kln}\left(\mathcal D_{a} K_{il}-2{ }^{(\gamma)}\pi^p_a K_{r(i}\epsilon_{l)pr} \right) \left(\mathcal D_{b} K_{jk}-2{ }^{(\gamma)}\pi^q_b K_{s(j}\epsilon_{k)qs} \right) \\
&+{ }^{(\gamma)}E^{-1}{ }^{(\gamma)}E^a_p { }^{(\gamma)}E^b_p { }^{(\gamma)}\left(\mathcal D_a K_{ij} - 2{ }^{(\gamma)}\pi_a^k K_{l(i}\epsilon_{j)kl}\right) \left( \mathcal D_b K_{ij} - 2{ }^{(\gamma)}\pi_b^m K_{n(i}\epsilon_{j)mn}\right) \\
&-{ }^{(\gamma)}E^{-1} { }^{(\gamma)}E^a_i { }^{(\gamma)}E^b_j \left(D_a K_{jl}-2{ }^{(\gamma)}\pi^k_a K_{m(j}\epsilon_{l)km}\right) \left(D_b K_{il}-2{ }^{(\gamma)}\pi^n_b K_{p(i}\epsilon_{l)np}\right).
\end{split} \end{equation}
Note that except for the Hamiltonian constraint, all of the rest constraints do not contain the parameter $\gamma$ explicitly. Hence $\gamma$ does not affect the gauge transformations they generate. However, the Hamiltonian constraint consists of 4 polynomials of $\gamma$ with different powers. This fact may lead to different dynamics for different values of $\gamma$ in the quantum theory.

The Poisson bracket between connection variable $A^i_a(x)$ and conformal constraint $\mathcal Q(\omega_{\ell})$ reflects the spatial conformal transformation of the connection variable. The conformal constraint reads
\begin{equation}
\mathcal Q(\omega_{\ell})=-\int_{\Sigma_t}d^3x\left[2(A^j_b-\Gamma^j_b){ }^{(\gamma)}E^b_j+K_{jl}\mathcal P^{jl}\right]\omega_{\ell}.
\end{equation}
Hence we have
\begin{equation} \label{p.b. A-Q}
\{A^i_a(x),\mathcal Q(\omega_{\ell})\}=-2\omega_{\ell}(x)[A^i_a(x)-\Gamma^i_a(x)]+\epsilon^{ijk}{ }^{(\gamma)}E^j_a { }^{(\gamma)}E^b_k \partial_b \omega_{\ell}(x).
\end{equation}

\subsection{The second scheme}
Unlike GR, Weyl gravity is conformally invariant. Eq.(\ref{c.t by Q in F.F.}) shows that the conformal transformations of the conjugate pair $\pi_a^i$ and $E^b_j$ admit the form in the canonical transformation in last subsection. Thus it is reasonable to consider the possibility that the canonical transformations with different values of $\gamma$ are actually conformally equivalent to each other. This is not the case for the canonical transformations defined in last subsection, since the other conjugate pair $K_{ij}$ and $\mathcal P^{kl}$ remains unchanged there while it should be changed by the conformal transformations. In fact, the conformally equivalent canonical transformations can be defined as
\begin{equation} \begin{aligned}
&\pi^i_a \to A^i_a=\Gamma^i_a+\gamma \pi^i_a, \\
&E^b_j \to \frac{1}{\gamma}E^b_j\equiv { }^{(\gamma)}E^b_j, \\
&K_{ij} \to \sqrt{\gamma}K_{ij} \equiv { }^{(\gamma)}K_{ij}, \\
&\mathcal P^{kl} \to \frac{1}{\sqrt{\gamma}}\mathcal P^{kl} \equiv { }^{(\gamma)}{\mathcal P}^{kl}.
\end{aligned} \end{equation}
Then the original geometric variables are related to the new variables by
\begin{equation} \begin{aligned}
h_{ab}&=\gamma { }^{(\gamma)} E { }^{(\gamma)} E^i_a { }^{(\gamma)} E^i_b, \\
\pi^{cd}&=\frac{1}{2\gamma { }^{(\gamma)} E}\left[{ }^{(\gamma)} E^{(c}_j { }^{(\gamma)} E^{d)}_i (A^i_a-\Gamma^i_a) { }^{(\gamma)} E^a_j - { }^{(\gamma)} E^c_j { }^{(\gamma)} E^d_j (A^i_a-\Gamma^i_a) { }^{(\gamma)} E^a_i-2{ }^{(\gamma)} K_{il}{ }^{(\gamma)}{\mathcal P}^{lj}{ }^{(\gamma)} E^{(c}_i { }^{(\gamma)} E^{d)}_j\right], \\
K_{ab}&=\gamma^{\frac12} { }^{(\gamma)} E { }^{(\gamma)} E^i_a { }^{(\gamma)} E^j_b { }^{(\gamma)} K_{ij}, \\
\mathcal P^{cd}&=\gamma^{-\frac12}{ }^{(\gamma)} E^{-1} { }^{(\gamma)} E^c_k { }^{(\gamma)} E^d_l{ }^{(\gamma)}{\mathcal P}^{kl}.
\end{aligned} \end{equation}
The constraints can be recast as
\begin{equation} \label{c.t. in c-d 2nd} \begin{aligned}
{\mathcal G}_i&=\mathcal D_a { }^{(\gamma)} E^a_i+2\epsilon_{ijk}{ }^{(\gamma)} K_{jl}{ }^{(\gamma)}{\mathcal P}^{lk}\approx 0, \\
{\mathcal P}&=\sqrt{\gamma}\delta_{ij}{ }^{(\gamma)}{\mathcal P}^{ij}\approx 0, \\
{\mathcal Q}&=-2(A^i_a-\Gamma^i_a){ }^{(\gamma)} E^a_i-{ }^{(\gamma)} K_{ij}{ }^{(\gamma)}{\mathcal P}^{ij}\approx 0, \\
{H}_a&=F^i_{ab} { }^{(\gamma)} E^b_i+{ }^{(\gamma)}{\mathcal P}^{ij} \mathcal D_a { }^{(\gamma)} K_{ij}-{ }^{(\gamma)} \pi^i_a{ }^{(\gamma)}{\mathcal G}_i\approx 0, \\
{H}_0&=\gamma^{-\frac12}\left({ }^{(\gamma)}{\mathcal H}_{A}+{ }^{(\gamma)}{\mathcal H}_{B}+{ }^{(\gamma)}{\mathcal H}_C+{ }^{(\gamma)}{\mathcal H}_D\right)\approx 0,
\end{aligned} \end{equation}
 where
\begin{equation} \begin{aligned}
{ }^{(\gamma)}\mathcal H_A=&-\frac{1}{2\sqrt{{ }^{(\gamma)} E}}{ }^{(\gamma)}\mathcal P^{ij}{ }^{(\gamma)}\mathcal P^{ij}, \\
{ }^{(\gamma)}\mathcal H_B=&\frac{1}{ { }^{(\gamma)} E}{ }^{(\gamma)} E^a_{(i} { }^{(\gamma)} E^b_{j)} \Big[ \mathcal D_a \mathcal D_b { }^{(\gamma)}\mathcal P^{ij}-4\epsilon^{i}_{\phantom{i}kl}{ }^{(\gamma)}\pi^k_a \mathcal D_b { }^{(\gamma)}\mathcal P^{jl} -2\epsilon^{i}_{\phantom{i}kl}  { }^{(\gamma)}\mathcal P^{jl}\mathcal D_b{ }^{(\gamma)}\pi^k_a \\
&+6{ }^{(\gamma)}\mathcal P^{jk}{ }^{(\gamma)}\pi^k_a { }^{(\gamma)}\pi^i_b-4{ }^{(\gamma)}\mathcal P^{ij}{ }^{(\gamma)}\pi^k_a{ }^{(\gamma)}\pi^k_b - 2\delta_{ij}{ }^{(\gamma)}\mathcal P^{kl}{ }^{(\gamma)}\pi^k_a{ }^{(\gamma)}\pi^l_b \Big]+\frac{1}{ { }^{(\gamma)} E}\mathcal P^{ij}{ }^{(\gamma)} E^a_j{ }^{(\gamma)} E^b_k R^{ik}_{ab}, \\
{ }^{(\gamma)}\mathcal H_C=&{ }^{(\gamma)}K_{ij}{ }^{(\gamma)}\pi^i_a { }^{(\gamma)} E^a_j-3{ }^{(\gamma)}K { }^{(\gamma)}\pi^i_a { }^{(\gamma)} E^a_i -2{ }^{(\gamma)}K_{ij}{ }^{(\gamma)}K_{il}{ }^{(\gamma)}\mathcal P^{lj}, \\
{ }^{(\gamma)}\mathcal H_D=&-\frac{1}{ \sqrt{{ }^{(\gamma)} E} }\Big[ { }^{(\gamma)}E^a_m { }^{(\gamma)}E^b_n \epsilon^{ijm}\epsilon^{kln}\left(\mathcal D_{a} { }^{(\gamma)}K_{il}-2{ }^{(\gamma)}\pi^p_a { }^{(\gamma)}K_{r(i}\epsilon_{l)pr} \right) \left(\mathcal D_{b} { }^{(\gamma)}K_{jk}-2{ }^{(\gamma)}\pi^q_b { }^{(\gamma)}K_{s(j}\epsilon_{k)qs} \right) \\
&+{ }^{(\gamma)}E^a_p { }^{(\gamma)}E^b_p { }^{(\gamma)}\left(\mathcal D_a { }^{(\gamma)}K_{ij} - 2{ }^{(\gamma)}\pi_a^k { }^{(\gamma)}K_{l(i}\epsilon_{j)kl}\right) \left( \mathcal D_b { }^{(\gamma)}K_{ij} - 2{ }^{(\gamma)}\pi_b^m { }^{(\gamma)}K_{n(i}\epsilon_{j)mn}\right) \\
&- { }^{(\gamma)}E^a_i { }^{(\gamma)}E^b_j \left(D_a { }^{(\gamma)}K_{jl}-2{ }^{(\gamma)}\pi^k_a { }^{(\gamma)}K_{m(j}\epsilon_{l)km}\right) \left(D_b { }^{(\gamma)}K_{il}-2{ }^{(\gamma)}\pi^n_b { }^{(\gamma)}K_{p(i}\epsilon_{l)np}\right)\Big].
\end{aligned} \end{equation}
Note that the Hamiltonian constraint in (\ref{c.t. in c-d 2nd}) consists of 4 terms of $\gamma$ with the same power.
In this connection-dynamical formalism, different values of the parameter $\gamma$ of the basic variables can be generated by particular conformal transformations. Since Weyl gravity is conformally invariant, the so-called Immirzi parameter ambiguity can be avoided in the corresponding loop quantum Weyl gravity. This observation can be confirmed by the fact that the parameter $\gamma$ can be removed from the expressions of all the constraints in (\ref{c.t. in c-d 2nd}).

\section{Summary}
In previous sections, the Hamiltonian structure of Weyl gravity has been studied in details. The conformal invariance of the theory is encoded in the conformal constraints $\mathcal Q(\omega_{\ell})$ and $\mathcal P(\omega_{\perp})$, which generate spatial and temporal conformal transformations respectively. The relation of the smeared fields $\omega_{\ell}$ and $\omega_{\perp}$ with the conformal factor $\Omega$ is worked out as (\ref{omega_ell}) and (\ref{omega_perp}). The Hamiltonian geometrodynamics of Weyl gravity is then recast into triad formalism by including the internal gauge degrees of freedom of a triad. The relation of the basic variables in triad formalism and the original ones is worked out as (\ref{Frame hKP}) and (\ref{pi}). The rotation constraint (\ref{rotation}) is imposed for recovering the phase space of geometrodynamics from the extended phase space. It is shown that the new constrained system is still first class as that in geometrodynamics. In comparison to the case of original phase space, the conformal transformations generated by $\mathcal P(\omega_{\perp})$ on the extended phase space take simpler forms. The variable $\pi_a^i$ conjugate to the densitized triad $E^b_j$ keeps unchanged by the temporal conformal transformations, and only the diagonal elements of the components of the extrinsic curvature $K_{ij}$ are affected by it.

The main purpose of this paper is to construct certain connection dynamical formalism of Weyl gravity, in order to apply the method of LQG to this theory. This purpose has been realized by two schemes of canonical transformations on the extended phase space. In the first scheme, only the conjugate pair $(\pi_a^i, E_j^b)$ are transformed into an $SU(2)$ connection and its momentum, while the other conjugate pair $(K_{ij}, \mathcal P^{kl})$ keep unchanged. The so-called Immirzi parameter $\gamma$ ambiguity in LQG of GR exists also in the corresponding quantum theory of Weyl gravity in this formalism. However, in the second scheme, both conjugate pairs are transformed, and the canonical transformations with different values of the parameter $\gamma$ are related by certain conformal transformations generated by the constraint $Q(\omega_{\ell})$. Therefore, the connection formalisms with different values of $\gamma$ belong to a conformally equivalent class. There will be no Immirzi parameter ambiguity in the corresponding quantum theory in this formalism. This intriguing feature of connection formalism of Weyl gravity deserves further investigating in its loop quantization. Another interesting issue in both schemes is the role played by the conjugate pair $(K_{ij}, \mathcal P^{kl})$ in the connection-dynamical formalism. From the expressions of the Gaussian constraint and diffeomorphism constraint in (\ref{c.t. in c-d}) or (\ref{c.t. in c-d 2nd}), $(K_{ij}, \mathcal P^{kl})$ or $({ }^{(\gamma)} K_{ij}, { }^{(\gamma)} \mathcal P^{kl})$ look like certain internal tensor valued matter fields in GR. This implies possible geometrical origin of certain matter fields from Weyl gravity, which also deserves further investigating in its quantum theory.

\section*{Acknowledgments}
This work is supported by the NSFC (Grant Nos. 11475023 and 11875006). Y. M. would like to thank Abhay Ashtekar for his generous
hospitality during his visit to Penn State where part of this work was done.

\appendix
\section{Conformal transform by assembled generator} \label{C=Q+P}
One can write down the 0th and first order terms of $\exp[\mathcal C(\omega_{\ell},\omega_{\perp})]K_{ab}$, and then iterate the procedure to obtain
\begin{equation} \begin{aligned}
\text{0th} \qquad \qquad &K_{ab}\\
\text{1st} \qquad \qquad &\omega_{\ell}K_{ab}+\omega_{\perp}h_{ab}\\
\text{2nd} \qquad \qquad &\omega_{\ell}^2 K_{ab} + 3\omega_{\ell}\omega_{\perp}h_{ab} \\
\text{3rd} \qquad \qquad &\omega_{\ell}^3 K_{ab} + 7\omega_{\ell}^2\omega_{\perp}h_{ab} \\
\text{4th} \qquad \qquad &\omega_{\ell}^4 K_{ab} + 15\omega_{\ell}^2\omega_{\perp}h_{ab} \\
&\cdots \\
n\text{th} \qquad \qquad & \omega_{\ell}^n K_{ab} + (2b_{n-1}+1)\omega_{\ell}^{n-1}\omega_{\perp}h_{ab} \\
(n+1)\text{th} \qquad \qquad & \omega_{\ell}^{(n+1)} K_{ab} + (2b_n+1)\omega_{\ell}^n\omega_{\perp}h_{ab} \\
\end{aligned} \end{equation}
Thus we have to solve the sequence $b_{n+1}=2b_n+1$ and get its solution as $b_n=2^n-1$. Therefore the Taylor series of $\exp[\mathcal C(\omega_{\ell},\omega_{\perp})]\circ K_{ab}$ are expressed by two equations
\begin{equation} \label{finite C=Q+P}
\left\{ \begin{aligned}
& \bar\Omega=\Omega\vert_{\Sigma_t}=\sum^{\infty}_{n}\frac{1}{n!}\omega_{\ell}^n=e^{\omega_{\ell}} \\
& n^{\mu}\partial_{\mu}\Omega=\omega_{\perp}\sum^{\infty}_{n=0}\frac{2^n-1}{n!}\omega_{\ell}^{(n-1)}
\end{aligned} \right.
\end{equation}

\end{document}